\documentstyle[preprint,aps,prb]{revtex} 
\begin{document}         
\draft  
\title{Influence of orbital pair breaking on 
         paramagnetically  limited states  
         in clean superconductors}
\author{ S.~Manalo and U.~Klein }
\address{Johannes Kepler Universit{\"a}t Linz, Institut f{\"u}r 
Theoretische Physik, A-4040 Linz, Austria}
\date{July 16, 2001}
\maketitle  
\begin{abstract}  
Paramagnetic pair breaking is believed to be of 
increasing importance in many layered superconducting 
materials such as cuprates and organic compounds.
Recently, strong evidence for a phase transition to 
the Fulde-Ferrell-Larkin-Ovchinnikov(FFLO)
state has been obtained for the first time. We present a 
new theory of competing spin and orbital pair breaking in 
clean superconducting films or layers. As a general 
result, we find that the influence of orbital 
pair breaking on the 
paramagnetically limited phase boundary is rather strong, 
and its neglect seldom justified. This is particularly 
true for the FFLO state which can be destroyed by a  
very small orbital contribution. We discuss the situation 
in $\mbox{Y}\mbox{Ba}_{2}\mbox{Cu}_{3}\mbox{O}_{7}$ which has 
two coupled conducting Cu-O layers per unit cell. As 
a consequence, an intrinsic orbital pair breaking 
component might exist even for applied field exactly
parallel to the layers.
\end{abstract} 

\section[intro]{Introduction}
\label{sec:intro}
Most theoretical studies of paramagnetic pair breaking 
in superconductors followed the attitude of the 
classical papers by Clogston \cite{CLOGSTON} and 
Chandrasekhar \cite{CHANDRA} where only
spin pair breaking was considered and the 
orbital component was assumed to be 
negligibly small. A notable exception is the dirty limit
theory developed by Maki \cite{MAKPREV}, Fulde 
\cite{FULDEREV} and others. In many experiments, on the 
other hand, {\em both\/} pair breaking components are present
and the neglect of the orbital contribution is not really 
justified. Recently, ultra-thin films became available and 
several new classes of layered 
superconducting compounds have been discovered. For 
applied field parallel to the films \cite{BUTADA}
or conducting planes \cite{DKCLBFSS}, 
Pauli paramagnetism can be the dominating pair 
breaking effect, provided the conducting layers are 
sufficiently separated from each other or the thickness 
of the films is sufficiently small.  In many of these 
compounds, including High-$\mbox{T}_{c}$ cuprates 
and organic superconductors, impurity scattering and 
spin-orbit coupling is small and orbital pair breaking is 
- for an applied field parallel to the planes - the most 
important second order effect, next to the spin effect, to 
be taken into account. 

Of particular interest is the 
Fulde-Ferrell-Larkin-Ovchinnikov(FFLO) 
state \cite{FUFE,LAROVCH}, 
which is a spatially inhomogeneous 
superconducting state, predicted to occur in 
clean superconductors with purely paramagnetic 
limiting.
Recent critical field measurements \cite{NSSBAP} in the   
quasi-two-dimensional organic superconductor 
$\kappa-\mbox{(BEDT-TTF)}_2\mbox{Cu(NCS)}_2$  strongly 
suggest that a state of the 
FFLO type exists  in this material; agreement 
between experiment \cite{NSSBAP} and existing 
theories has been successfully checked 
\cite{MANKL} both in view of the 
angle-dependence \cite{SHIMARAIN} and the 
temperature dependence \cite{MAKWON} of the 
upper critical field (see also \cite{SSNAKD}).  
Apparently, this is the first time since the 
original predictions in 
1964 \cite{FUFE,LAROVCH} that quantitative agreement 
between theory and experiment with regard to 
the FFLO phase boundary has been established.
Strong paramagnetic effects can also be expected for the 
High-$\mbox{T}_c$  cuprate superconductors at low 
temperatures, when the conducting planes in adjacent unit 
cells are well separated from each other.  A measurement 
\cite{DKCLBFSS,BNDCKLSMMGRB} at $T=1.6\,\mbox{K}$ in 
$\mbox{Y}\mbox{Ba}_{2}\mbox{Cu}_{3}\mbox{O}_{7}$ indicates 
rather clearly that the superconducting state is 
paramagnetically limited but, on the other hand, the 
observed transition is too broad to allow a decision  
between the FFLO state and the homogeneous 
superconducting state. 

A measure of the relative strength of orbital 
and paramagnetic pair breaking is the 
ratio of the paramagnetic critical field $H_p$ 
divided by the orbital upper critical 
field $H_{c2}$ of a type II superconductor. 
For a bulk superconductor in the clean limit this ratio 
can be written as 
\begin{equation}
  \label{eq:relpauliupper}
  H_p/H_{c2}  \sim \frac{\xi_0}{k_F^{-1}}
\mbox{,}
\end{equation}  
in terms of the Fermi wavelength $k_F$ and the coherence 
length $\xi_0$ of BCS theory.
 
This relation implies that orbital pair breaking
will always be the dominating mechanism in bulk 
superconductors, no matter how large the 
Ginzburg-Landau (GL) parameter $\kappa$ 
is; this holds at least in the 
framework of conventional BCS theory. In a thin 
superconducting layer of thickness $d<\xi_0$ 
, on the other hand, the orbital critical field 
$H_{c}(d)$ is increased by a factor of $\xi_0/d$ and 
the corresponding ratio is given by 
\begin{equation}
  \label{eq:relpaulifilm}
  H_p/H_{c}(d)  \sim \frac{d}{k_F^{-1}}
\mbox{.} 
\end{equation}
In comparison to Eq.\ (\ref{eq:relpauliupper})  
a small transverse dimension 
$d\ll\xi_0$ of the film suppresses 
the orbital effect and 
enlarges the spin effect 
drastically.  However, equation  
(\ref{eq:relpaulifilm}) also shows, that
the critical thickness which separates the 
spin pair-breaking and 
orbital pair-breaking dominated regimes is 
still of the order of an atomic distance.
Thus, the estimate  
(\ref{eq:relpaulifilm}), which is confirmed by 
more quantitative calculations to be 
presented below, indicates that a nearly 
perfect two-dimensional situation
is required in order to justify the 
neglect of orbital pair breaking 
contribution in clean 
superconductors.
(The situation in dirty 
superconductors is much more favorable for 
the spin effect; the FFLO state, however,
is suppressed by impurities).  

The simultaneous action of both types of pair breaking 
has already been studied for a particular situation, 
an infinitely thin superconducting film in a tilted 
magnetic field \cite{BULATILT,SHIMARAIN,KLRAISHI}. In 
such a configuration, orbital pair breaking is 
entirely due to the perpendicular field component, 
while the component parallel to the film is 
exclusively responsible for the spin
effect. The states near the upper critical field 
can be characterized by different Landau quantum 
numbers $n=0,1,2,\dots$, depending on the tilt 
angle \cite{BULATILT,SHIMARAIN}. The order parameter 
structure near $H_{c2}$ has also been investigated and 
shows interesting properties, such as several  
zeros with different vorticity per unit cell 
\cite{KLRAISHI}. For exactly plane-parallel external 
field the FFLO phase is recovered. However, in this 
limit no orbital pair breaking exists and 
only the spin effect survives, as a consequence of
the vanishing thickness of the superconducting 
layer in this model. 

In this paper we investigate a superconducting 
film of {\em finite thickness\/} in a magnetic field parallel
to the conducting plane. Thus, the usual model 
of purely paramagnetic pair breaking is generalized
in a  different way, taking into account the 
influence of a finite orbital pair breaking component 
on the FFLO state. The model is formulated in section 
\ref{sec:baseq}, using the framework of the quasiclassical 
Eilenberger equations. We assume that the film thickness 
is smaller than the coherence length and use a cylindrical 
Fermi surface. This shape of the Fermi surface allows 
us to study the influence of orbital pair breaking 
without any additional complications, like scattering 
of quasiparticles at the film boundaries. Such boundary 
effects seem less important in the present context, 
but may, nevertheless, be present in many materials 
and should be taken into account in future work. In 
addition, the cylindrical shape of the Fermi surface, 
which corresponds to a truly two-dimensional situation, 
allows us to extend our investigations to superconducting 
layers of atomic dimension. The main results, obtained by 
solving numerically the relevant phase boundary and 
stability equations, and a discussion of possible
orbital pair breaking contributions in the plane-parallel 
field configuration of 
$\mbox{Y}\mbox{Ba}_{2}\mbox{Cu}_{3}\mbox{O}_{7}$, are 
reported in section \ref{sec:resdis}.   Finally, the results 
are summarized in section \ref{sec:summarize}.

\section[baseq]{Basic Equations}  
\label{sec:baseq}
We first calculate the highest field where a superconducting 
solution of the quasiclassical equations, for small order 
parameter $\Delta$, exists in a thin film. This field may 
correspond to a second order phase transition or to the 
supercooling limit of the normal conducting state; to 
make a decision between these two possibilities 
the free energy of the competing homogeneous 
superconducting state will be calculated in a 
second step. 

\subsection[stablim]{Stability limit of normal conducting state}
\label{sec:Stab-limit-norm}

Let the film be parallel to the $xy$ plane with a finite 
extension from $-d/2$ to $+d/2$ in the $z-$direction.  
The applied magnetic field $B$ is  assumed to be parallel to
the plane of the film and parallel to the $y-$direction, 
$\vec{B}=B\vec{e}_y$. The transport equations, linearized 
in $\Delta$, are given by 
\begin{equation}
  \label{eq:quasif}
\left[ 2(\omega_l-\imath \mu B) +\hbar \vec{v}_F(\hat{k})\left( \vec{\nabla}_r-
\imath(2e/\hbar c) \vec{A}\right) \right] 
f(\vec{r},\hat{k},\omega_s) = 2\Delta(\vec{r},\hat{k}) \mbox{,}
\end{equation}
\begin{equation}
  \label{eq:quasifplus}
\left[ 2(\omega_l-\imath \mu B)- \hbar \vec{v}_F(\hat{k}) 
\left(
\vec{\nabla}_r+\imath(2e/\hbar c) \vec{A}      
\right)
 \right] 
f^{\it+}(\vec{r},\hat{k},\omega_s) = 2\Delta^\ast(\vec{r},\hat{k}) 
\mbox{,}
\end{equation} 
Here, the Zeeman term $\mu B$ occurs in the combination 
$\omega_s=\omega_l-\imath \mu B$, where $\omega_l=(2l+1)\pi k_BT$ are Matsubara frequencies, 
$\mu\simeq\hbar |e|/(2mc)$ is the magnetic moment of the electron 
and $B$ is the magnitude of the induction. The 
self-consistency equation for the gap is given by
\begin{equation}
  \label{eq:scop}
\left(2 \pi k_B T  \sum_{l=0}^{N_D} \frac{1}{\omega_l}+ 
\ln \frac{T}{T_c} \right) 
\Delta(\vec{r},\hat{k}) = \pi k_B T  \sum_{l=0}^{N_D} 
\int d^2 \hat{k}^\prime \,V(\hat{k},\hat{k}^\prime) \,
\left[ f(\vec{r},\hat{k}^\prime,\omega_s) +
f(\vec{r},\hat{k}^\prime,\omega_s^\ast ) \right] 
\mbox{,}
\end{equation}
where $N_D$ is the cutoff index for the Matsubara sums.  
The Fermi velocity is given by $\vec{v}_F(\hat{k})=
v_F(\vec{e}_x\cos\varphi +\vec{e}_y\sin{\varphi})= v_F\hat{k}$; the integral in 
Eq.\ (\ref{eq:scop}) over the cylindrical Fermi 
surface is simply a one-dimensional integral 
over the angle variable $\varphi$. We allow for a separable 
gap anisotropy, $\Delta(\vec{r},\hat{k})=
\Delta(\vec{r})\gamma(\hat{k}),\,
V(\hat{k},\hat{k}^\prime)
=\gamma(\hat{k})\gamma(\hat{k}^\prime)$, which 
will be specialized later to s-wave and d-wave 
superconductivity. We use the following 
gauge for the vector potential:  
$A_x=Bz\mbox{, and } A_y=A_z=0$. 

The standard method to solve the linearized 
transport equations uses a complete set of 
eigenfunctions of the operator 
$\hat{k}\vec{\partial_r}$ to construct 
the inverse of the 
differential operators on the l.h.s. of 
Eqs.\ (\ref{eq:quasif}),\,(\ref{eq:quasifplus}).  
Here, $\vec{\partial_r}$ 
is an abbreviation for the gauge-invariant 
derivative, $\vec{\partial_r}=\vec{\nabla}_r-
\imath(2e/\hbar c) \vec{A}$. For a 
cylindrical Fermi surface, $\hat{k}\vec{\partial_r}$ 
contains no derivative with respect to $z$ and 
the Green's functions depend on $z$ in a purely 
local way ($z$ playing the role of a parameter). 
This allows a straightforward generalization of 
the standard method to the present problem.      

Let us start from the well-known bulk solution of 
eqs.~(\ref{eq:quasif}),(\ref{eq:quasifplus}). 
If the eigenfunctions and eigenvalues of the 
operator $\hat{k}\vec{\partial_r}$ for an 
infinite sample are denoted by 
$f_{\hat{k}\vec{p}}$ and $\imath \hat{k}\vec{p}$ 
respectively,   
\begin{equation}
  \label{eq:eigenproblem}
\hat{k}\vec{\partial_r}\,f_{\hat{k},\vec{p}}(\vec{r})=
\imath \hat{k} \vec{p}\,f_{\hat{k},\vec{p}}(\vec{r})
\mbox{,}
\end{equation}
then the solution of Eq.\ (\ref{eq:quasif}) 
is given by  
\begin{equation}
  \label{eq:soloffgreen}
f(\vec{r},\hat{k},\omega_s)=
\int {\rm d}^3r_1
\int \frac{{\rm d}^3p}{(2\pi)^3}
\frac{f^*_{\hat{k},\vec{p}}(\vec{r_1}) 
f_{\hat{k},\vec{p}}(\vec{r})
}{\omega_s + \imath v_F \hat{k}\vec{p}/2}\,
\Delta(\vec{r}_1,\hat{k})
\mbox{,}
\end{equation}
where the spatial integration extends over all space.
In the chosen gauge the solutions of 
Eq.\ (\ref{eq:eigenproblem}) are given by 
\begin{equation} 
  \label{eq:eigensolapp} 
f_{\hat{k},\vec{p}}(\vec{r})=
{\rm e}^{
-\frac{\imath}{2}
\left[
\frac{2|e|}{\hbar c} (\hat{k}\vec{r}) 
\left(\left(\vec{B}\times\vec{r}\right)\hat{k} \right)+
\kappa_{\parallel}zx
\right]
+\imath\vec{p}\vec{r}
}
\mbox{,}
\end{equation} 
with the abbreviation $\kappa_{\parallel}=
\frac{2|e|}{\hbar c}B$. Using the   
completeness of the set of eigenfunctions 
(\ref{eq:eigensolapp}) the Green's functions  
may immediately be written in the form of 
Eq.\ (\ref{eq:soloffgreen}).  

To transform relation (\ref{eq:soloffgreen}) to 
a finite volume, it is, in our case, only necessary 
to restrict the spatial integration in 
Eq.\ (\ref{eq:soloffgreen}) to the 
film volume, i.e. to perform the integration 
over $z_1$ from $-d/2$ to $+d/2$. This simple 
method works only for a cylindrical Fermi 
surface, where the momentum of the quasiparticles 
is always parallel to the film boundaries. 
Otherwise, quasiparticle scattering at the film 
boundaries leads, for small $d<\xi_0$, to a 
modification of the integral kernel which has to 
be calculated by solving Eq.\ (\ref{eq:eigenproblem})
in a finite volume, with appropriate boundary 
conditions.

To proceed, the denominator of the integrand in 
Eq.\ (\ref{eq:soloffgreen}) is shifted into an 
argument of an exponential function by means of the 
identity 
\begin{equation}
  \label{eq:identity}
\frac{1}{r}=\int_{0}^{\infty}{\rm d}t\,{\rm e}^{-rt} 
\mbox{.}
\end{equation}
Now, if the eigenfunctions (\ref{eq:eigensolapp}) are 
inserted, the Green's function (\ref{eq:soloffgreen}) is 
represented as an integral with regard to the variables  
$\vec{p},\,\vec{r},\,\mbox{and }t$ over an exponential 
function. Two of these integrations can be performed 
analytically and the Green's function takes the form    
\begin{equation}
  \label{eq:aintresult}
f(\vec{r},\hat{k},\omega_s)=\int_{0}^{\infty}{\rm d}t\,
{\rm e}^{-t\omega_s}{\rm e}^
{
-\frac{\imath}{2}
t v_F \kappa_{\parallel}z \hat{k}_x 
}\Delta (\vec{r}-\frac{t}{2}v_F\hat{k},\hat{k})
\mbox{.}
\end{equation}

For $d<\xi$ the order parameter may be considered as 
$z$-independent and the Greens's function 
$f(\vec{r},\hat{k},\omega_s)$ may be replaced by its  
value $f(\hat{r},\hat{k},\omega_s)$ which depends 
only on $x$ and $y$ and denotes the average 
of $f(\vec{r},\hat{k},\omega_s)$, 
with respect to $z$, from $-d/2$ to $+d/2$: 
\begin{eqnarray}
  \label{eq:favereagerint}
f(\hat{r},\hat{k},\omega_s)=\int_{0}^{\infty}{\rm d}t\,
{\rm e}^{-t\omega_s}
\frac{1}
{\frac{1}{4}t v_F \kappa_{\parallel}d \hat{k}_x}
\sin 
\left( \frac{1}{4}t v_F \kappa_{\parallel}d \hat{k}_x \right)
\Delta (\hat{r}-\frac{t}{2}v_F\hat{k},\hat{k})
\mbox{.} 
\end{eqnarray}
The rest of the calculation is a straightforward 
generalization of methods developed in previous   
works \cite{LAROVCH,SHIMARAIN}. It is
convenient to perform the following shift in the 
argument of the space-dependent part of the gap:
\begin{equation}
  \label{eq:shiftdeltaarg}
\Delta (\hat{r}-\frac{t}{2}v_F\hat{k})=
{\rm e}^{-\frac{1}{2}t v_F \hat{k}\vec{\partial_r}}
\Delta (\hat{r})
\mbox{.}
\end{equation}

Inserting the Green's function solutions in the 
self-consistency equation for the gap (\ref{eq:scop}) and 
performing the Matsubara sum yields the linearized 
gap equation 
\begin{eqnarray}
  \label{eq:lingapsenkparal}
-\ln \left(\frac{T}{T_c}\right)\Delta(\hat{r}) &=&
\pi k_B T \int_{0}^{\infty}{\rm d}t \frac{1}
{\sinh(\pi T t)} \int_{0}^{2\pi}
\frac{{\rm d} \varphi^{\prime}}
{2\pi}\,\gamma( \hat{k}^{\prime})^{2} 
\cdot \bigg[1-
\frac{1}
{\frac{1}{4}t v_F \kappa_{\parallel}d {\hat{k}_x}^{\prime}}
\nonumber\\ & & \cdot
\sin 
\left( \frac{1}{4}t v_F \kappa_{\parallel}d 
{\hat{k}_x}^\prime \right) 
 \cos \left(\mu Bt-\frac{1}{2\imath} tv_F 
\hat{k}^{\prime}\hat{\nabla} 
\right) 
\bigg]
\Delta(\hat{r})
\mbox{}   
\end{eqnarray}
which has to be solved in order to find the magnetic 
field where the normal-conducting state breaks down.
The operator $\hat{\nabla}$ used in Eq.\ (\ref{eq:lingapsenkparal})
acts in the $x,y$-plane. 
Eq.\ (\ref{eq:lingapsenkparal}) differs 
from previous results \cite{SHIMARAIN} by a $d$-dependent 
factor which reduces to $1$ in the limit $d\to0$. For d-wave
superconductivity the finite thickness of the film breaks
rotational invariance; if $\Phi$ is the angle between the 
magnetic field and the $y-$axis of the crystal, the 
following replacement has to be performed in the integrand
of Eq.\ (\ref{eq:lingapsenkparal}):
  \begin{equation}
  \label{eq:directiondep}
d \hat{k}_x^\prime \Rightarrow 
d\left(
\hat{k}_x^\prime \cos \Phi -\hat{k}_y^\prime \sin \Phi 
 \right)
\mbox{,}
\end{equation}
in order to take the angle dependence of the external 
field into account.

To proceed further, the gap $\Delta$ is assumed 
to be proportional to plane wave states 
${\rm e}^{\imath\hat{q}\hat{r}}$. Solving the linearized 
gap equation\ (\ref{eq:lingapsenkparal}), with $\hat{\nabla}$ 
replaced by $\imath \hat{q}$,  
for different wave numbers $\hat{q}$, one obtains a function
$B(\hat{q})$. The field we are looking for, where the normal 
conducting state breaks down - and the corresponding wave 
number - is given by the highest $B(\hat{q})$. 
 
Eq.\ (\ref{eq:lingapsenkparal}) comprises two  
limiting cases where the behavior of the solutions is
known, the purely paramagnetic limit (for $d=0$),
and the purely orbital limit (for $\mu=0$). For 
$d=0$ one obtains the standard FFLO result \cite{FUFE},
\cite{LAROVCH}. For $\mu=0$ 
Eq.\ (\ref{eq:lingapsenkparal}) may be solved 
analytically near $T_c$. In this limit one obtains for the 
parallel critical field, where the normal-conducting 
solution breaks down
\begin{equation}
  \label{eq:glresult}
B^{\parallel}_{c}=0.61\frac{h c}{2 e}
\frac{\sqrt{1-T/T_{c}}}{\xi_{0}d}
\mbox{.}
\end{equation}    
Eq.\ (\ref{eq:glresult}) is in agreement with the second 
order transition line obtained for $d<\xi(T)$ in the 
framework of GL theory. At low temperatures 
microscopic calculations have, to our knowledge, only 
be performed for Fermi surfaces of spherical 
\cite{SHAPOVAL,USADEL,SCOPESCH} or ellipsoidal \cite{SCOPESCH}
shape. In these works a $d^{-3/2}$ behavior for the 
critical field has been obtained 
in the limit of very small $d$. In contrast, the present 
theory, using a cylindrical Fermi surface, leads to an 
approximate $d^{-1}$ behavior of the critical field in the 
whole temperature range.

\subsection[phasebound]{Phase boundaries of homogeneous 
states} 
\label{sec:Phase-bound-homog}
For an infinitely thin film, the free energy of the 
FFLO state is only slightly lower than the free energy 
of the homogeneous (paramagnetically limited) 
superconducting state. The presence of an orbital  
pair breaking component, realized by a vector 
potential, in our film of thickness $d$ may change 
the free energy balance in a decisive way.
To clarify this point, we calculate the free energies of 
the homogeneous superconducting and normal-conducting 
states and compare the resulting phase boundaries and 
stability limits with the FFLO transition line studied 
in the last subsection. Considering films or layers of 
finite thickness, we refer to states which do 
not depend on the coordinates $x,y$ within the plane 
as `homogeneous states`; these states may, nevertheless, 
depend on $z$ as a consequence of a (residual) screening 
property of the thin film.  

The spatially constant, purely paramagnetically limited 
superconducting state has been studied 
first by Sarma \cite{SARMA}; this case corresponds to $d=0$ in
the present model. For $T/T_c<0.56$ he found a first order 
phase boundary, which lies below the FFLO transition line.
This line is determined by inserting the solutions of the 
nonlinear gap equation
\begin{equation}
  \label{eq:scophom}
2 \pi k_B T  \sum_{l=0}^{N_D} \frac{1}{\omega_l}+ 
\ln \frac{T}{T_c} 
= \pi k_B T  \sum_{l=0}^{N_D} \,
\left( 
\frac{1}{\sqrt{|\Delta|^2+\omega_s^2}}+cc.
\right) 
\mbox{,}
\end{equation} 
into the free energy difference
\begin{equation}
  \label{eq:freeenhomogen}   
F_s-F_n=-N(E_F)\pi k_B T\sum_{l=0}^{N_D}
\left(
\frac{\left(\omega_s-\sqrt{|\Delta|^2+\omega_s^2} 
\right)^{2}}
{\sqrt{|\Delta|^2+\omega_s^2}}+cc.
\right)
\mbox{.}
\end{equation}
For $T<0.56\,T_c$ the gap as a function of $B$ 
has two branches \cite{SARMA,BURKRAIN}, as shown in 
Fig.\ \ref{fig:stablimdiszero}. 
Our calculation of the second variation of the 
free energy shows (see the curves for $T/T_c=0.1$ 
in Fig.\ \ref{fig:stablimdiszero}) that the homogeneous 
superconducting state may be superheated up to the 
highest field, where the two branches cross. 
The lowest field where the lower branch exists 
defines, on the other hand, the supercooling limit  
of the normal-conducting state. Thus, the region of 
the lower branch corresponds, as expected,  exactly to 
the metastable region of the first order transition.
%
%
In this way, three transition lines, the phase 
transition line where the free energies coincide,  
the superheating line, and the supercooling line, are 
determined by solving 
Eqs.\ (\ref{eq:scophom}),\,(\ref{eq:freeenhomogen}); 
for finite $d$ the same method is used to determine 
the metastable region.

For the present circular Fermi surface, the determination 
of the 'homogeneous states' in a film of finite thickness 
$d$ is still a local problem despite the 
nontrivial $z-$dependence appearing in the transport 
equations. The assumption of a gap which depends weakly 
on the $z-$ coordinate leads, in analogy to the 
reasoning of the last subsection, to the following 
relation between the averaged Green's functions 
$\bar{f},\,\bar{g}$ and the gap $\Delta$:
\begin{equation}
  \label{eq:greenrelsmalld}
\bar{f}=\frac{1}{B(d,\varphi)}
\arctan (\frac{B(d,\varphi)}{\omega_s})\Delta \bar{g}
\mbox{,}
\end{equation}
where 
\begin{equation}
  \label{eq:defbvondphi}
B(d,\varphi)=\frac{v_F\kappa_{\parallel} d \cos \varphi }
{4\pi k_B T_c}
\mbox{.}
\end{equation}
Note, that orbital pair-breaking leads to a dependence of
the Green's functions on the quasiparticle 
wave number $\hat{k}$. 
Using Eq.\ (\ref{eq:greenrelsmalld}) the self-consistency 
relation for the gap takes the form
\begin{equation}
  \label{eq:scophomfind}
2 \pi k_B T  \sum_{l=0}^{N_D} \frac{1}{\omega_l}+ 
\ln \frac{T}{T_c} 
= \pi k_B T  \sum_{l=0}^{N_D} 
\int_0^{2\pi}\frac{d\varphi}{2\pi}
\left( 
\frac{\beta(\varphi)}{\sqrt{|\Delta|^2\beta(\varphi)
+A_l(d,\varphi)^2}}+cc.
\right) 
\mbox{,} 
\end{equation}
and the free energy difference is given by 
\begin{equation}   
  \label{eq:freeenhomogenfind} 
F_s-F_n=-N(E_F)\pi k_B T\sum_{l=0}^{N_D}
\int_0^{2\pi}\frac{d\varphi}{2\pi}
\left(
\frac{\left(A_l(d,\varphi)-\sqrt{|\Delta|^2\beta(\varphi)+
A_l(d,\varphi)^2} \right)^{2}}
{\sqrt{|\Delta|^2\beta(\varphi)+A_l(d,\varphi)^2}}+cc.
\right)
\mbox{,} 
\end{equation}
where the factor $A_l(d,\varphi)$ is defined by   
\begin{equation}
  \label{eq:defofafactor}  
\frac{1}{A_l(d,\varphi)}=\frac{1}{B(d,\varphi)}
\arctan (\frac{B(d,\varphi)}{\omega_s})
\mbox{.} 
\end{equation}
The factor $\beta(\varphi)$ is $1$ for s-wave and 
$1+\cos(4\varphi)$ for d-wave superconductivity. 
Eqs.\ (\ref{eq:scophomfind}),\,(\ref{eq:freeenhomogenfind})
are essentially of the same (local) form as 
Eqs.\ (\ref{eq:scophom}),\,(\ref{eq:freeenhomogen}); the 
main difference is the dependence of the Green's functions
on the direction $\varphi$ of the quasiparticle momentum, 
which leads to the $\varphi-$integrals in 
Eqs.\ (\ref{eq:scophomfind}),\,(\ref{eq:freeenhomogenfind}). 
For $d \to 0$ the previous results are recovered as $A_l(d,\varphi)$ 
approaches $\omega_s$ in this limit. Solving 
Eq.\ (\ref{eq:scophomfind})
and calculating the free energy 
\ (\ref{eq:freeenhomogenfind}) the influence of a finite 
orbital pair breaking contribution, due to a 
nonzero film thickness $d$, on the three transition 
lines can be studied.    
\section[resdis]{Results and Discussion}
\label{sec:resdis}
In this section we discuss the following four transition 
lines, defined in more detail in the last section: (1) 
the line $B_c$ where the free energies of the homogeneous 
superconducting and normal-conducting states coincide, 
(2) the superheating limit $B_{sh}$ of the homogeneous 
superconducting state, (3) the supercooling limit 
(stability limit) of the normal-conducting state against 
spatially homogeneous superconducting fluctuations, which 
is denoted by $B_{sc}$, and (4) the stability limit of the 
normal-conducting state against spatially inhomogeneous 
superconducting fluctuations, denoted by $B_{FFLO}$. The 
detailed results reported here are mainly for isotropic 
gap (s-wave) superconductors; a few 
calculations have also been performed for d-wave 
superconductors in view of recent experiments on 
$\mbox{Y}\mbox{Ba}_{2}\mbox{Cu}_{3}\mbox{O}_{7}$. 

As a starting point, we show in 
Fig.\ \ref{fig:overviewfordiszero} these four transition 
lines for $d=0$, in the purely paramagnetic limit.  
For $T>T_{tri}=0.56\,T_c$ all four lines merge into a single 
second order transition line. Below the tricritical 
point $T_{tri}$ only three different lines are visible in 
Fig.\ \ref{fig:overviewfordiszero} since, interestingly, 
$B_{FFLO}$ and $B_{sh}$ exactly coincide (this coincidence 
occurs, however, only for a circular Fermi surface). 
Only $B_{FFLO}$ is physically significant for $d=0$, the 
other lines are meaningless. At $B_{FFLO}$ a second order 
phase transition to the FFLO state takes place; the 
lower phase boundary of the FFLO state 
is not dealt with here; it has been studied by Burkhardt 
and Rainer \cite{BURKRAIN}.

The paramagnetic pair-breaking effect 
dominates for very small $d$ while the orbital effect 
dominates for large $d$. Thus, it should be possible to
define a critical thickness $d^{\ast}$ which roughly separates 
the two regimes. This crossover behavior is shown in 
Fig.\ \ref{fig:crossoverbehavior} for the thermodynamic 
critical field $B_c$. The value of $d^{\ast}$ is
of the order of $k_F^{-1}$ in the region of low $T$, 
in agreement with the estimate of section~\ref{sec:intro}. 
Fig.\ \ref{fig:crossoverbehavior} shows also the decreasing 
importance of paramagnetic pair breaking with increasing 
$T$.  
Generally, the additional orbital pair-breaking effect 
brought about by the finite thickness of the conducting 
layer, leads to a depression of all four fields shown 
in the reference figure (Fig.\ \ref{fig:overviewfordiszero}). 
A detailed plot of the $T-$dependence of the fields 
$B_{FFLO},\,B_c,\,B_{sc}$ (the superheating 
field $B_{sh}$ lies above $B_{FFLO}$ and has been 
omitted for clarity) for $d/k_F^{-1}=0.5,\,1.0,\,2.5$ 
as shown in Fig.\ \ref{fig:btplan3fieds} reveals, however, 
significant differences. The FFLO transition field is much 
stronger suppressed than the line $B_c$ where the free 
energies of the homogeneous states coincide. As a 
consequence, for $d\gtrsim 0.5\,k_F^{-1}$ (see top of 
Fig.\ \ref{fig:btplan3fieds}) the FFLO 
transition vanishes ($B_{FFLO}$ becomes the supercooling 
field of the normal state) and is replaced by a first 
order transition at $B_c$ to the homogeneous 
superconducting state. Note that 
a conducting layer of atomic thickness (dimension of 
unit cell in the plane) yields enough 
orbital pair breaking to produce this suppression of 
$B_{FFLO}$ in favor of $B_c$. This behavior is not 
unreasonable; spatially varying states are known to be 
much more sensitive to perturbations than homogeneous 
states (recall in this context Andersons theorem). 
With further increasing orbital pair breaking (see 
middle and bottom of Fig.\ \ref{fig:btplan3fieds})
the lines $B_{FFLO}$ and $B_{sc}$ tend to merge and the 
metastable region shrinks; for $d>3\,k_F^{-1}$ orbital 
pair-breaking dominates.

\begin{figure}[htbp]
  \begin{center}
    \caption{Critical fields 
$B_{FFLO},\,B_c,\,B_{sc},\,B_{sh}$
as a function of $d$ for $T=0.01\,T_c$.}
    \label{fig:bdplan4fields}  
  \end{center}
\end{figure}
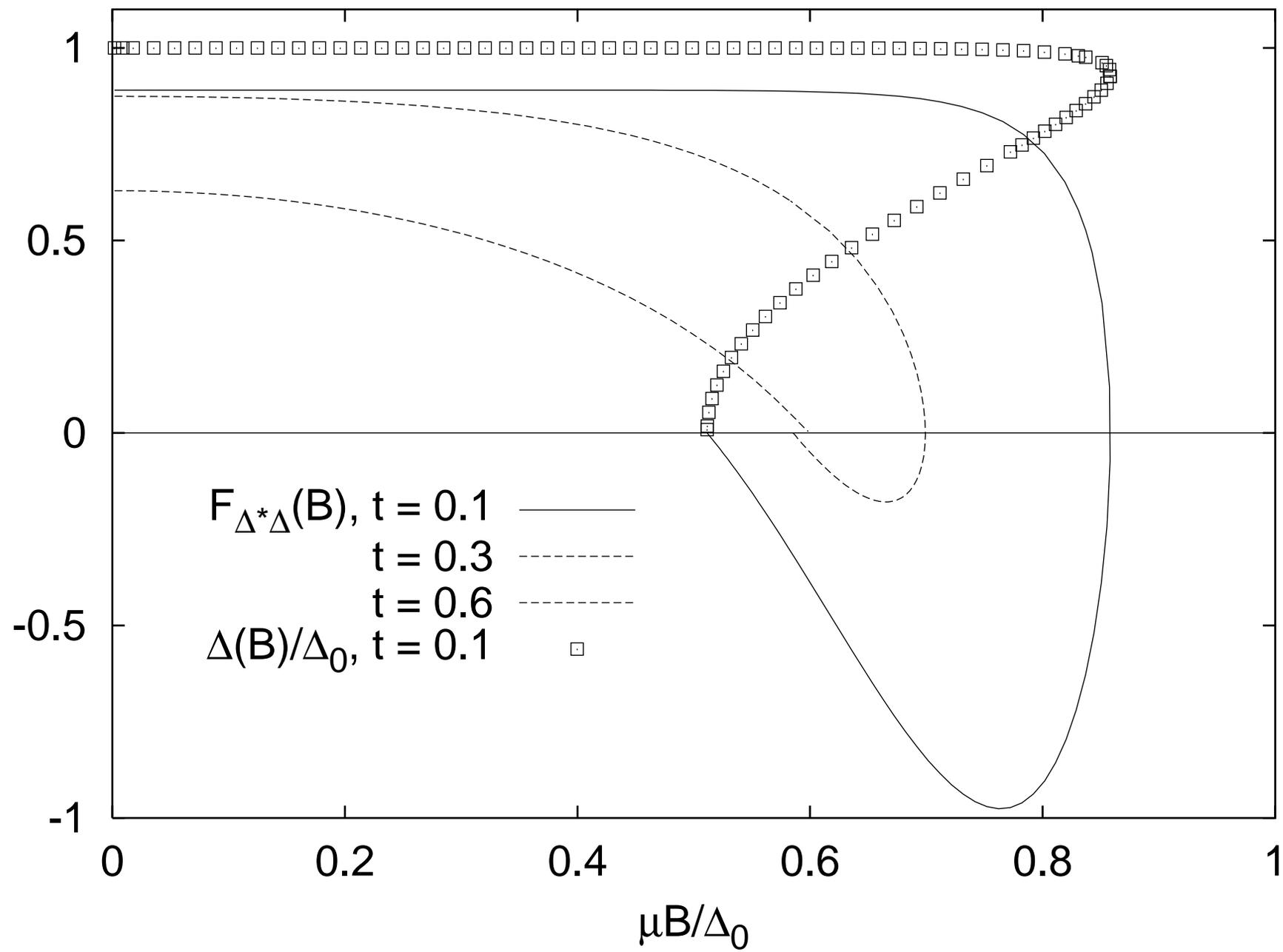
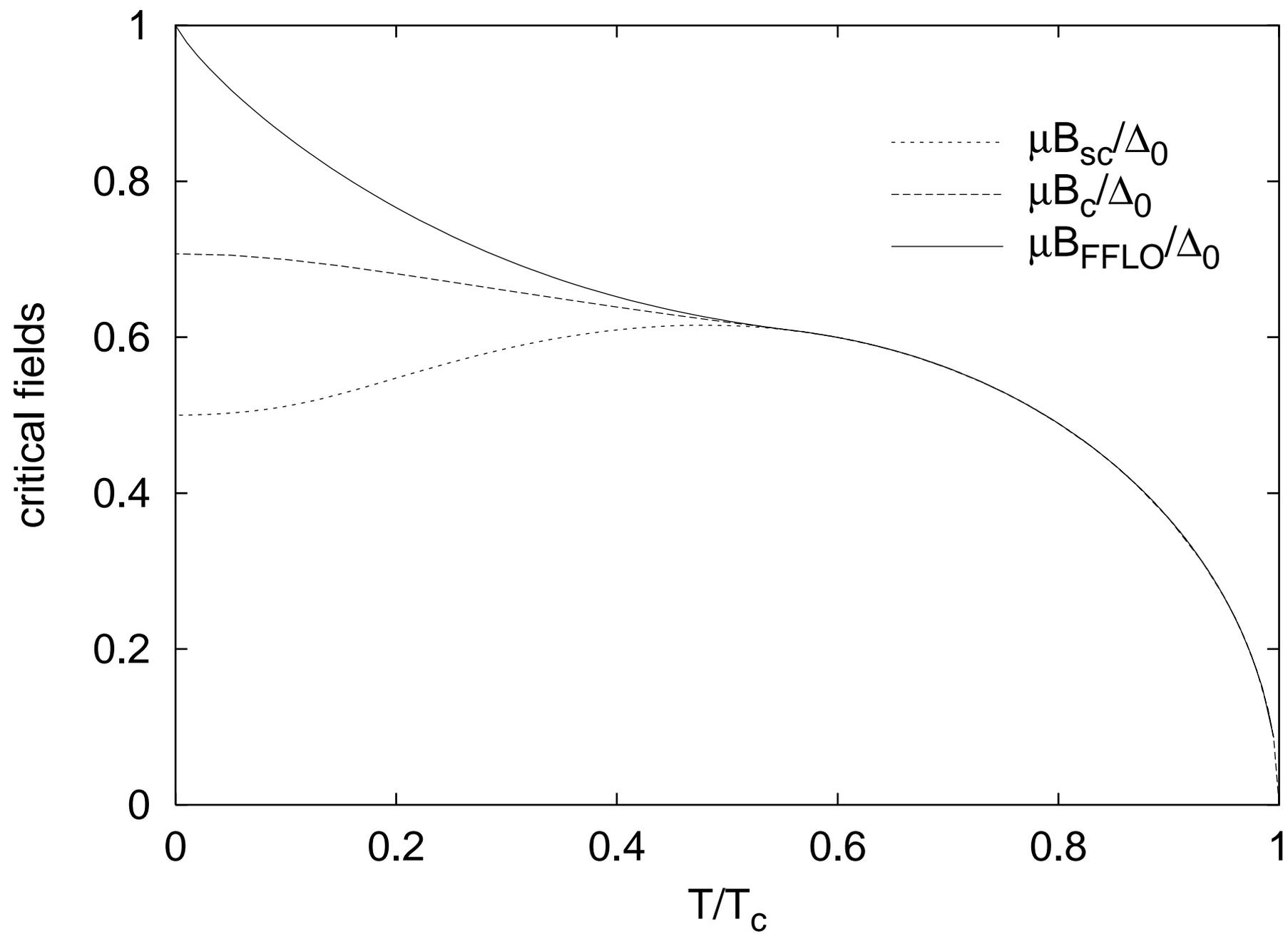
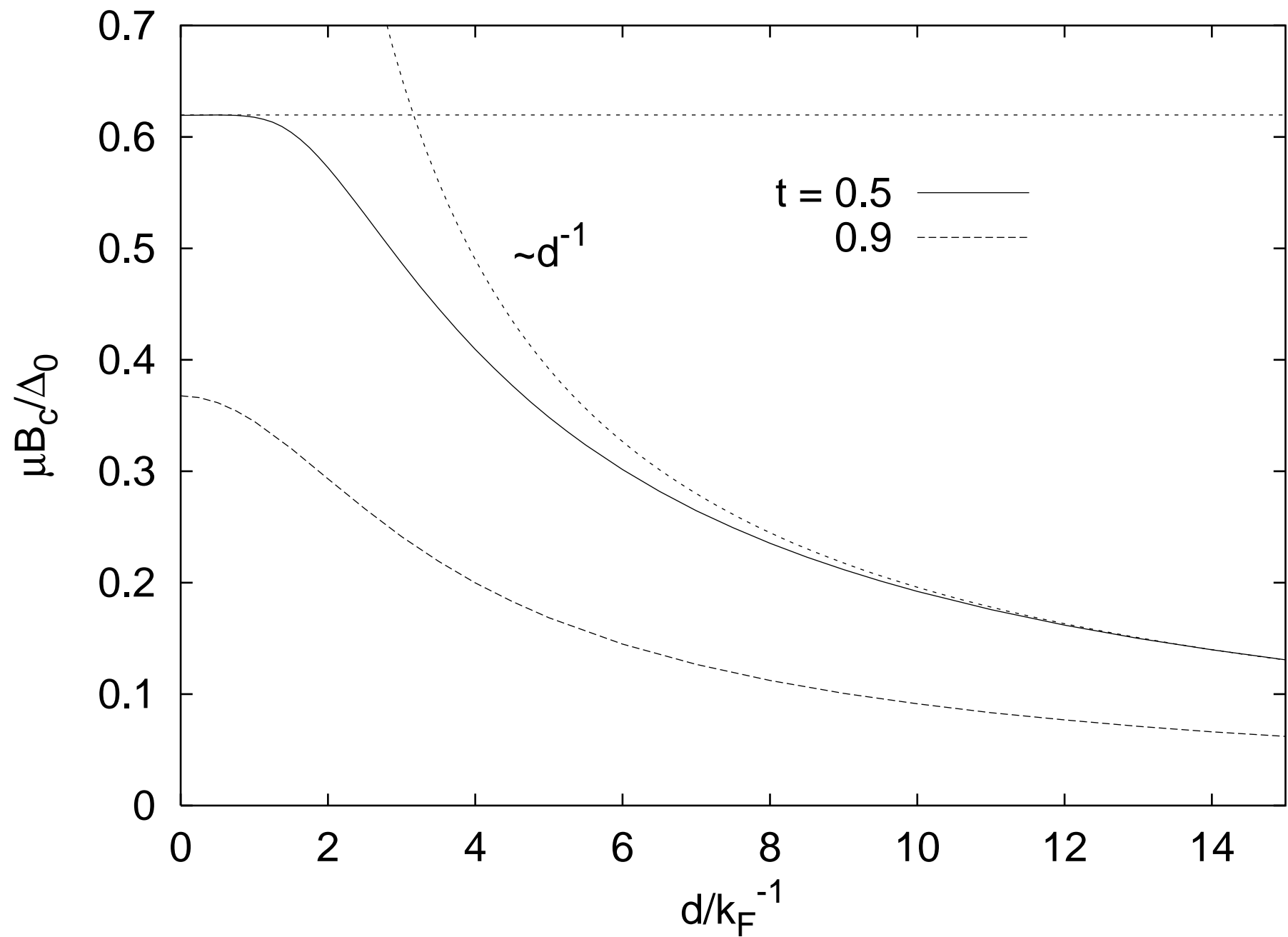
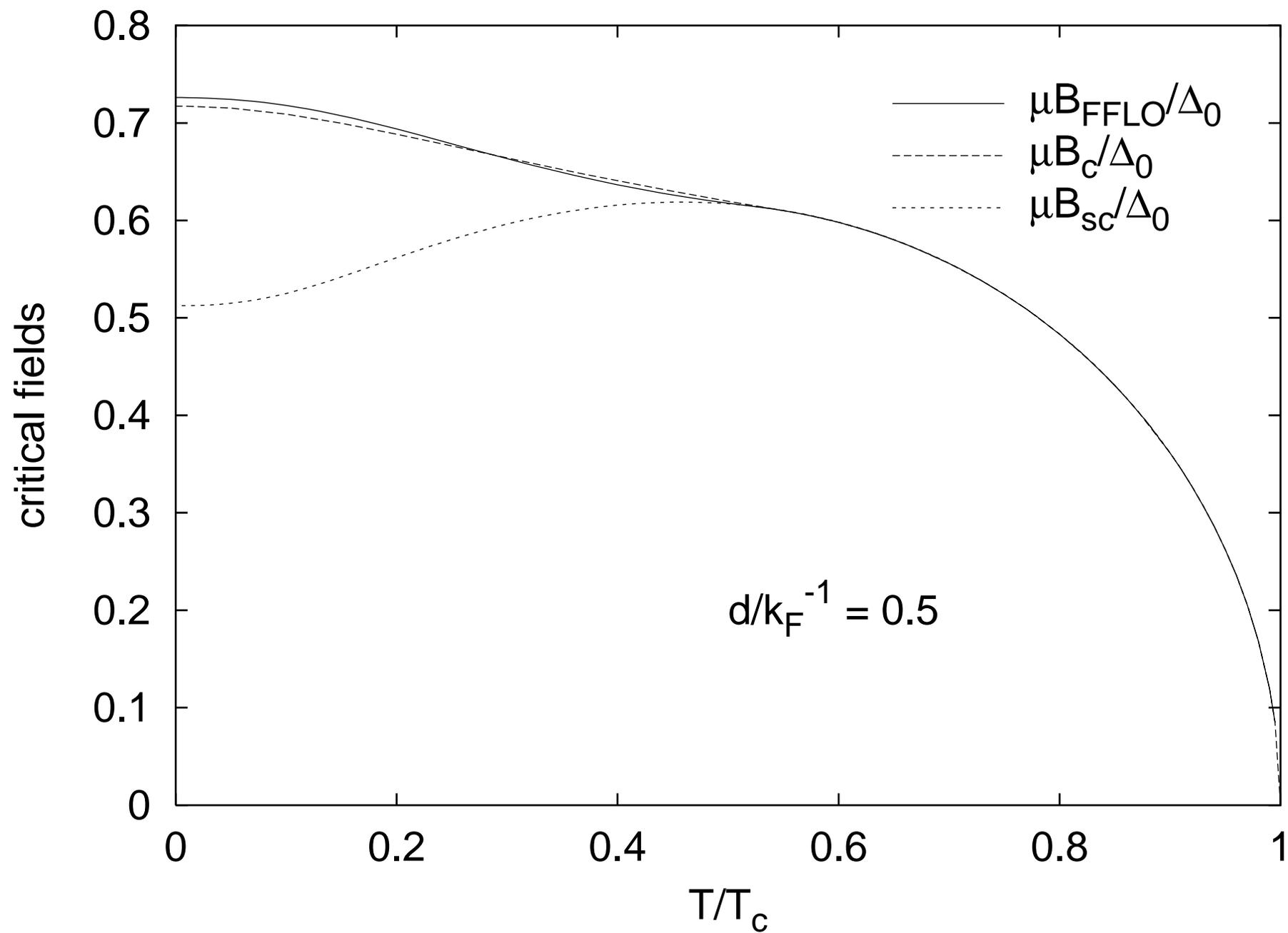
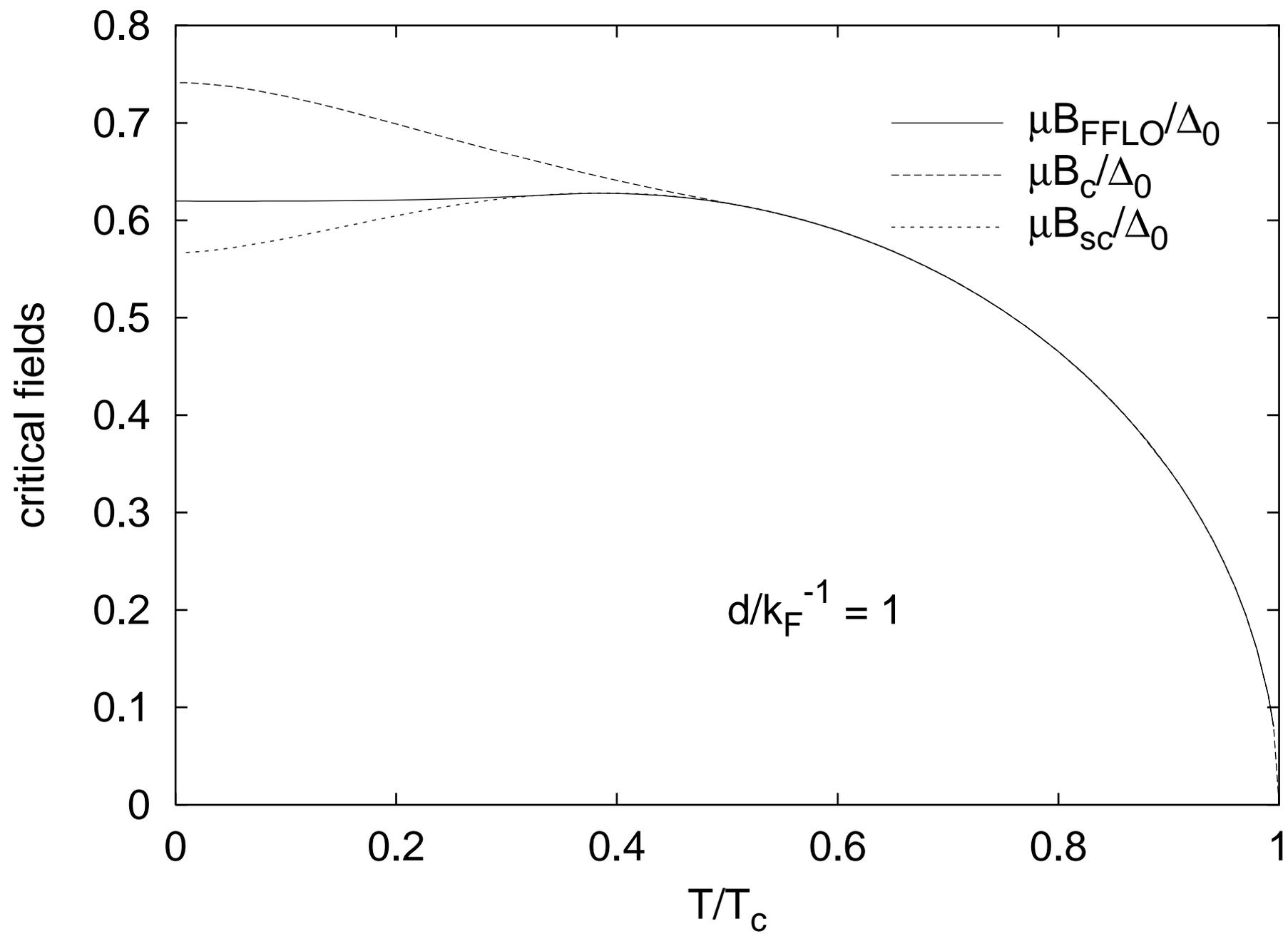
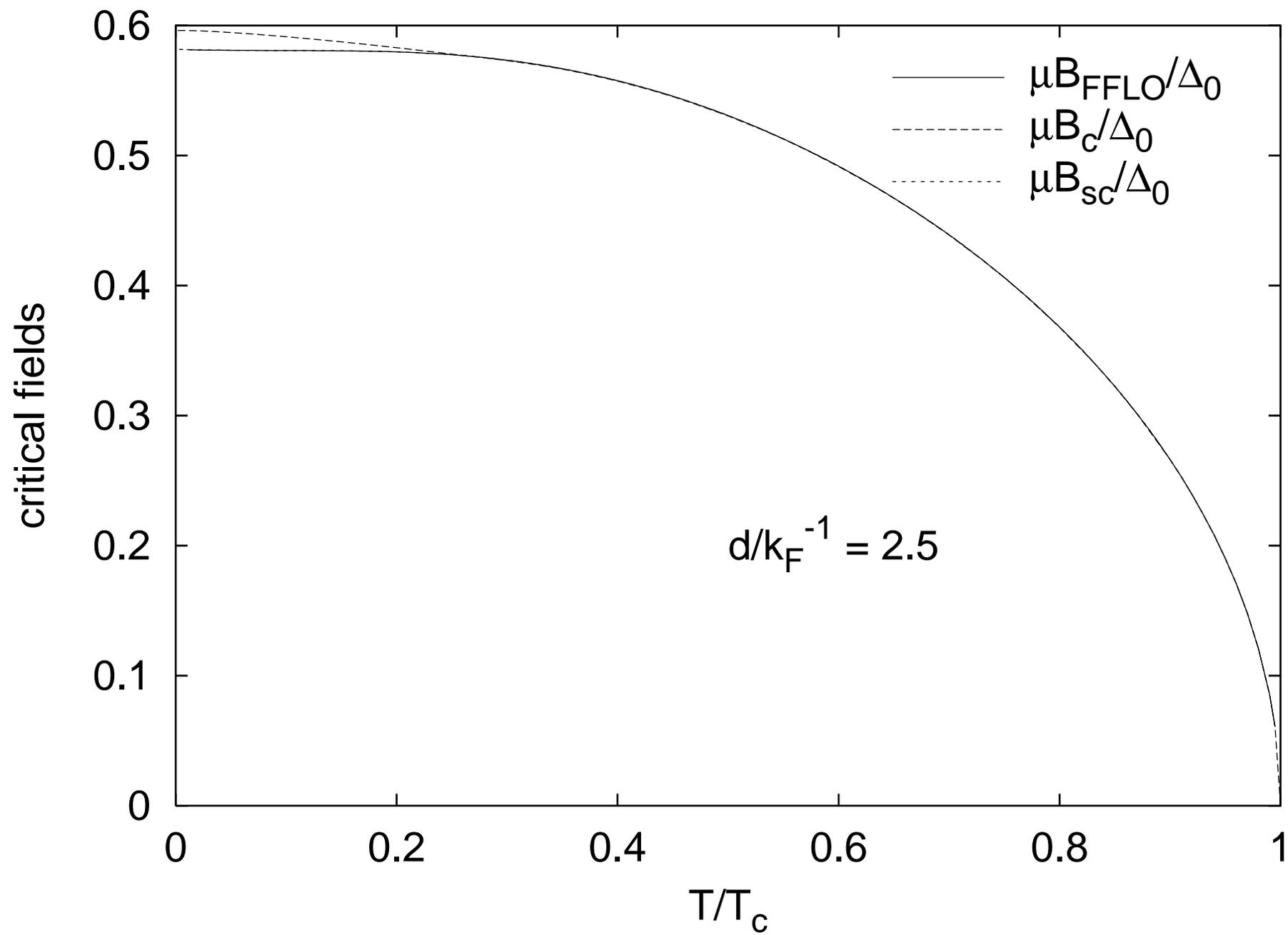
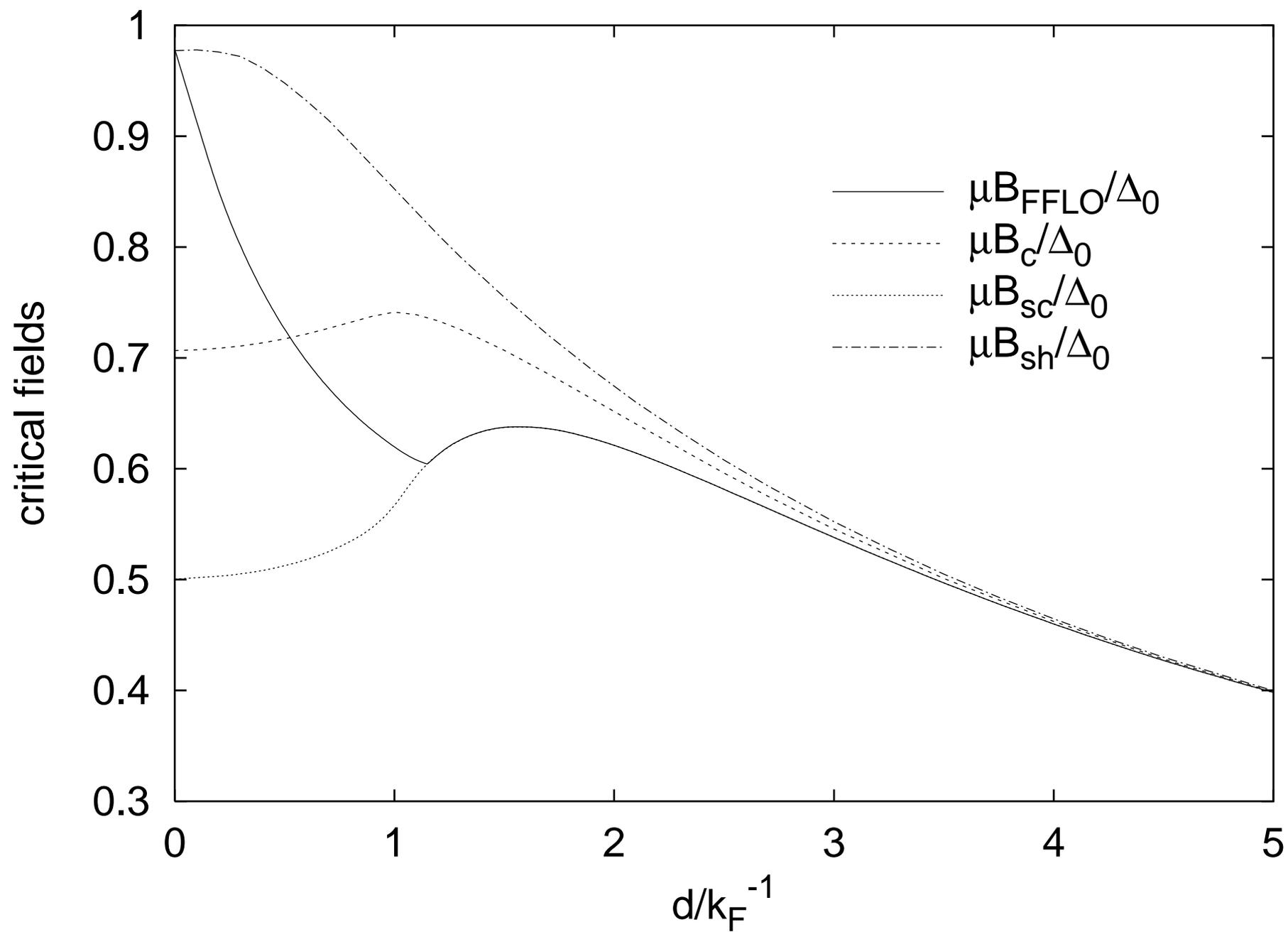

The behavior of all four fields as a function of $d$ 
is shown in Fig.\ \ref{fig:bdplan4fields} in the 
low temperature region (for $T=0.01\,T_c$), 
where paramagnetic effects are most pronounced.
This figure gives an overview of the cross-over from the 
paramagnetically  dominated regime at very small $d$ to the 
orbitally dominated regime at large $d$. If the thickness 
where $B_{FFLO}$ and $B_c$ cross is denoted by $d_1$, then
the FFLO state is only realized in the small range 
$d<d_1 \cong 0.5\,k_F^{-1}$, for $d>d_1$ a first order transition to 
the homogeneous (mainly) paramagnetically limited 
state occurs. The FFLO line plays the role of a supercooling
limit of the normal state until it falls (at 
$d \cong 1.2 k_F^{-1}$) below the line $B_{sc}$, where the 
normal-conducting state is limited by spatially constant 
superconducting fluctuations. The wavenumber $q$ of the 
FFLO state decreases with increasing $d$ until 
it jumps to $q=0$ at the line $B_{sc}$ (at the 
crossing point two degenerate solutions exist for $q$). 
It should be 
pointed out that we continue to use here the term 'FFLO 
state', even if this term denotes, strictly speaking, a 
state without any orbital pair breaking contribution.
With increasing $d$ the lines $B_{sh},\,B_c,\,B_{sc}$ 
approach each other and the transition becomes 
identical to the well-known second order transition 
of a thin film in a parallel field, which is entirely 
due to orbital pair breaking (the region of really 
large $d$ where the difference between type I and 
type II superconductivity becomes important is clearly 
outside the range of validity of the present model). 

The results reported so far have been restricted to 
s-wave superconductors, with isotropic gap 
[using $\gamma(\hat{k})=1$ in  
Eqs.\ (\ref{eq:lingapsenkparal}),\,(\ref{eq:scophomfind})] 
. A few calculations have also been performed
for d-wave superconductors, using $\gamma(\hat{k})=
2(\hat{k}_x^{2}-\hat{k}_y^{2})$. The results generally 
confirm the behavior found for s-wave superconductors.
An interesting peculiarity of d-wave superconductors 
without any orbital pair-breaking is a steep rise 
of $B_{FFLO}$ with decreasing $T$ below $T/T_c =0.1$ 
(see Fig.\ 4 of Manalo and Klein \cite{MANKL}). This peak 
belongs to  
the $\phi_q=0$ portion of the critical field \cite{MAKWON}
and is much steeper than the corresponding part of 
the critical field curve for s-wave superconductors. Our 
calculations show that this peak can be effectively 
suppressed by a very small ($d/k_F^{-1} \approx 0.2$) 
amount of orbital pair breaking. The absence of this peak 
in measurements \cite{NSSBAP} on 
$\kappa-\mbox{(BEDT-TTF)}_2\mbox{Cu(NCS)}_2$ may be an 
indication of a very small residual orbital pair-breaking 
contribution in this material; a more detailed study would 
be worth while.
 
Let us investigate the consequences of the fact 
that the FFLO state can be suppressed in favor of 
the homogeneous superconducting state by a very small 
admixture of orbital pair-breaking (see Figure 
\ref{fig:bdplan4fields}). We have not mapped out a complete
phase diagram like figure \ref{fig:bdplan4fields} for d-wave 
superconductors (where $B_{FFLO}$ becomes anisotropic as 
a consequence of the finite $d$) but instead performed a 
few calculations in order to get an overview of what happens.
The results confirm qualitatively the main feature 
visible in figure \ref{fig:bdplan4fields}, namely a much 
stronger suppression of $B_{FFLO}$, as compared to 
$B_c$, by orbital pair-breaking. For 
$\mbox{Y}\mbox{Ba}_{2}\mbox{Cu}_{3}\mbox{O}_{7}$ we have two 
conducting $\mbox{CuO}_{2}$ layers per unit cell with a 
distance of $\approx 3.9\,{\AA}$, while the c-axis 
zero-temperature coherence length is estimated \cite{WKCVL} 
to be $2-4\,{\AA}$. The coupling between 
bilayers in adjacent unit cells may obviously be neglected, 
as a consequence of the large length $c\approx11.7\,{\AA}$ 
of the unit cell in this direction. The coupling between 
the two layers in one unit cell, on the other hand, 
remains an open question, and the following two 
possibilities should be taken into consideration. 

The first possibility is, that the two 
superconducting layers decouple at low $T$, below some 
crossover temperature $T^{\ast}$. As is well known, the 
orbital 
critical field of weakly coupled layers 
diverges \cite{KLJUBE} below some crossover 
temperature $T^{\ast}$, which means 
that paramagnetic pair-breaking is the only remaining 
mechanism to limit the superconducting state.
This requires a two-dimensional, in-plane mechanism of 
superconductivity. The amount of orbital pair-breaking 
would be negligibly small in this case and the 
superconducting state below the critical field should be 
the FFLO state. 

The second possibility is that the superconducting 
state keeps its finite extension for arbitrary $T$. This 
requires an inter-plane
mechanism where the bilayer structure is essential for 
the superconducting pairing process. In this case, the 
bilayer may be approximately replaced by a 
single layer of finite thickness $d\approx2-4\,{\AA}$ . 
Taking a 
value of $v_F\approx 7\cdot 10^{7} \mbox{cm}/\mbox{sec}$ for 
the Fermi velocity in the $a-b$ plane, as measured by
Andreev reflections \cite{HIDDMW}, we estimate a value 
between $1$ and $4$ for our dimensionless thickness 
parameter $d/k_F^{-1}$, which measures the amount of 
orbital pair-breaking. Thus in this case, if the 
bilayer structure can be approximated by a finite slab,  
the amount of intrinsic orbital pair-breaking 
in YBCO, brought about by the finite thickness of this 
slab, will be by far large enough to suppress the FFLO state.
The second order FFLO transiton will be replaced by a 
first order transition to a homogeneous 
superconducting state; this transition is 
due to the combined action of {\em both\/} 
pair breaking mechanisms rather than a single one.
For a typical value
of $d/k_F^{-1}\approx 2$, the critical field would be still of the 
same order of magnitude as the purely paramagnetically 
limited, (Pauli limiting) field at $d=0$ but with a 
strongly reduced metastability region 
(see Fig.\ \ref{fig:bdplan4fields})
 
The transport measurement of Dzurak et al. \cite{DKCLBFSS} 
of the critical field of 
$\mbox{Y}\mbox{Ba}_{2}\mbox{Cu}_{3}\mbox{O}_{7}$ at $1.6 \mbox{K}$
led to  a result of the order of the Pauli limiting field 
$B_p=B_c(d=0)$; the transition seems, however, too broad 
to distinguish between the FFLO and Pauli limiting fields. 
Further, more accurate experiments are required to 
settle this question, which concerns fundamental aspects 
of the superconducting state in High-$\mbox{T}_{c}$ cuprates. 
Observation of the FFLO state in the plane-parallel field 
configuration of $\mbox{Y}\mbox{Ba}_{2}\mbox{Cu}_{3}\mbox{O}_{7}$ 
would be a strong argument in favor of an 
in-plane mechanism of superconductivity in this material. 
The relevance of this question has also been 
discussed by Yang and Sondhi \cite{YANSON2}, using the 
framework of the Lawrence-Doniach model, which is in 
a sense complementary to the present approach.  

The orbital pair-breaking effect due to the finite 
thickness of the  conducting layers has previously 
been taken into account in a theory by Schneider and 
Schmidt \cite{SCHSCH}. This theory may be used 
successfully to fit the experimental 
data \cite{BNDCKLSMMGRB} near $T_c$, but neglects all  
paramagnetic effects. The purely paramagnetic limit, 
on the other hand, has been studied by Maki and 
Won \cite{MAKWON} and by Yang and Sondhi \cite{YANSON}.
The present results show that {\em both\/} effects should be 
taken into account for a detailed description of the 
transition. The orbital effect cannot be neglected 
even if the orbital critical field is several times 
higher than the paramagnetic limiting field. 

Finally, we note, that single atomic (molecular) layers 
are responsible for the superconducting state in the organic 
compound $\kappa-\mbox{(BEDT-TTF)}_2\mbox{Cu(NCS)}_2$ where 
phase boundaries compatible with a d-wave version of the 
FFLO state have recently been observed. Thus, a 
considerable influence of orbital pair-breaking, which 
would suppress the FFLO state in favor of the homogeneous 
superconducting state, does not exist in this material. 
Measurements of the detailed temperature-dependence of 
$B_{FFLO}$ close to $T=0$ could reveal even small admixtures 
of an orbital pair-breaking component. 

\section[summarize]{Conclusion}
We developed a theory of competing paramagnetic and 
orbital pair-breaking effects in clean superconducting 
films and layers. The destructive influence of 
orbital pair-breaking on the FFLO state turned out to 
be stronger than commonly expected. It is necessary to 
have single-atomic layers in exactly plane-parallel fields 
in order to be able to neglect completely the orbital 
component and observe the ``pure'' FFLO state. We 
calculated only 
the upper phase boundary of the ``mixed'' FFLO state 
(which occurs in films of finite thickness as a 
consequence of the combined action of both pair breaking
mechanisms). The equilibrium structure as well as the 
lower transition line to the homogeneous superconducting 
state, have not yet been calculated for this 
new inhomogeneous state. Another challenging open question
is a microscopic treatment of Josephson coupling between 
layers of finite thickness. Our results show that a careful 
measurement of the plane-parallel critical field in 
$\mbox{Y}\mbox{Ba}_{2}\mbox{Cu}_{3}\mbox{O}_{7}$ could
give important information on the question whether an 
in-plane or an inter-plane mechanism is responsible for  
superconductivity in $\mbox{High-T}_c$ superconductors.     
\label{sec:summarize}
\section*{Acknowledgments}
We acknowledge stimulating discussions with D.Rainer, 
Bayreuth. This work was supported by the Austrian 
``Fonds zur F{\"o}rderung der wissenschaftlichen Forschung''
under Project P13552-TPH.   
\begin{figure}[htbp]
  \begin{center}  
    \caption{Stable and unstable branch of $\Delta$ as
a function of $B$ for $t=T/T_c=0.1$ and second 
derivative of free energy $F_{\Delta^{\ast}\Delta}$ 
for $t=0.1,\,0.3,\,0.6$.}
    \label{fig:stablimdiszero} 
  \end{center}
\end{figure} 
\begin{figure}[htbp]
  \begin{center}
    \caption{Critical fields $B_{FFLO}$, $B_c$, and $B_{sc}$ 
as a function of $T$ for $d=0$. The superheating field 
$B_{sh}$ agrees with $B_{FFLO}$ for the present  circular 
Fermi surface.}
    \label{fig:overviewfordiszero} 
  \end{center} 
\end{figure}
\begin{figure}[htbp]
  \begin{center}
    \caption{Critical field $B_c$ as a function of the 
dimensionless thickness parameter $k_F^{-1}/d$ for 
$t=T/T_c=0.5$ and $t=0.9$. The purely orbital 
pairbreaking curve, which is proportional to $d^{-1}$ 
and the purely paramagnetic curve, which is independent 
of $d$ are shown as dotted lines ($t=0.5$).}
    \label{fig:crossoverbehavior} 
  \end{center}
\end{figure}
\begin{figure}[htbp] 
  \begin{center}
%
%
    \caption{Critical fields $B_{FFLO},\,B_c,\,B_{sc}$
as a function of $T$ for $d/k_F^{-1}=0.5$ (top), 
$d/k_F^{-1}=1.0$ (middle), und $d/k_F^{-1}=2.5$ (bottom).} 
    \label{fig:btplan3fieds} 
  \end{center}
\end{figure}
\end{document}